\documentclass[twocolumn,nofootinbib,amsmath,amssymb,aps,prd,balancelastpage,superscriptaddress]{revtex4-1}

\usepackage{color}
\usepackage[active]{srcltx}
\usepackage{amsmath,amsfonts,amssymb,amsthm,amstext,amscd,eucal,srcltx}
\usepackage{epsfig,graphicx,bm}
\usepackage{epstopdf, epsf}
\usepackage{dcolumn}
\usepackage{hyperref}

\newcommand{\be}{\begin{equation}}
\newcommand{\ee}{\end{equation}}

\newcommand{\bse}{\begin{subequations}}
\newcommand{\ese}{\end{subequations}}
\newcommand{\bea}{\begin{eqnarray}}
\newcommand{\eea}{\end{eqnarray}}
\newcommand{\ba}{\begin{array}}
\newcommand{\ea}{\end{array}}
\newcommand{\bc}{\begin{center}}
\newcommand{\ec}{\end{center}}

\begin{document}
\preprint{IPM/P-2012/009}  
\vspace*{3mm}

\title{Hidden Non-Locality and Self-Superrenormalization of Quantum Gravity}%

\author{Andrea Addazi}
\email{andrea.addazi@lngs.infn.it}
\affiliation{Center for Theoretical Physics, College of Physics Science and Technology, Sichuan University, 610065 Chengdu, China}
\affiliation{INFN sezione Roma {\it Tor Vergata}, I-00133 Rome, Italy}

\begin{abstract}
\noindent

We show that the formation/evaporation of Black Holes (BH) unitarizes quantum gravity 
at all the orders of the perturbation theory.  Non-perturbative quantum effects 
save the scattering amplitudes from any polynomial divergences. 
Such a phenomena is intimately related to the dynamical emergence of an effective non-locality
as well as emergent modifications of the Heisenberg's uncertainty principle.
The BH production de-localizes quantum gravity vertices and propagators as a consequence of its holographically stored entropy.
In this sense, quantum gravity is a superrenormalizable theory, although non-locality is hidden in its action. 

\end{abstract}

\maketitle


{\it Introduction: entropy is the key.} Apparently, quantum gravity leads to violations of unitarity
if the transferred energies among scattering particles exceed the Planck scale. 
Indeed, quantum radiative corrections do not provide for any way out from these issues. 
This is a clear consequence of the fact that the gravitational coupling scales as 
\begin{equation}
\label{alpha}
\alpha_{G}=G_{N}E_{CM}^{2}\, ,
\end{equation}
where $E_{CM}$ is the Center of Mass energy and $G_{N}$ the Newton constant. 
However,  possible puzzles arrise from considering inclusions of non-perturbative effects.
Indeed, for super-planckian Center of Mass energies and for impact parameters that are 
around the Schwarzschild radius $R_{S}$ as
\begin{equation}
\label{superPl}
E_{CM}\geq  M_{Pl}\,\,\,\, b\leq R_{S}=2G_{N}E_{CM}\, , 
\end{equation}
the formation/evaporation of the Black hole (BH) is expected. 
Considering the BH formation process, we would expect a geometric cross section:
\begin{equation}
\label{cro}
\sigma_{G}=\pi R_{S}^{2}= 4\pi (G_{N}E_{CM})^{2}\, .
\end{equation}
This because the BH is considered as a completely 
absorptive Black disk\footnote{ Indeed, not so dissimilarly to color glass condensate approaches around the high energy gluon saturation.}.  
In other words, in the scattering framework of the wave partial decomposition, this will correspond 
to a critical inelasticity including the maximal angular momenta eigenvalue as follows: 
\begin{equation}
\label{IMA}
{\rm Im}A(J=bE_{CM}, E_{CM})\sim G_{N}E_{CM}^{2}\, . 
\end{equation}

The problem emerging here is that Eqs.[\ref{cro},\ref{IMA}] seem to violate unitarity for $E_{CM}>M_{PL}$,
i.e. the inclusion of BHs and other possible non-perturbative effects do not seem to solve the unitarity issue. 
This argument is generically accepted as a further motivation towards any searches of a UV completion of quantum gravity. 

On the other hand, a possible dissonance  with respect to this conclusion is that the geometric cross-section 
does not only grow with the energy but also with the BH radius and, therefore, with the distance. 
This fact was envisaged, as a possible way out from the UV divergences, by {\it Banks} \cite{B1,B2} and elaborated later on by {\it Dvali} and {\it Gomez}
\cite{Dvali:2010bf,Dvali:2010jz,Dvali:2011th} -- related to the BH N-portrait \cite{Dvali:2011aa,Dvali:2012gb,Dvali:2012rt}.

\begin{figure}[ht]
\centerline{ \includegraphics [width=1.1\columnwidth]{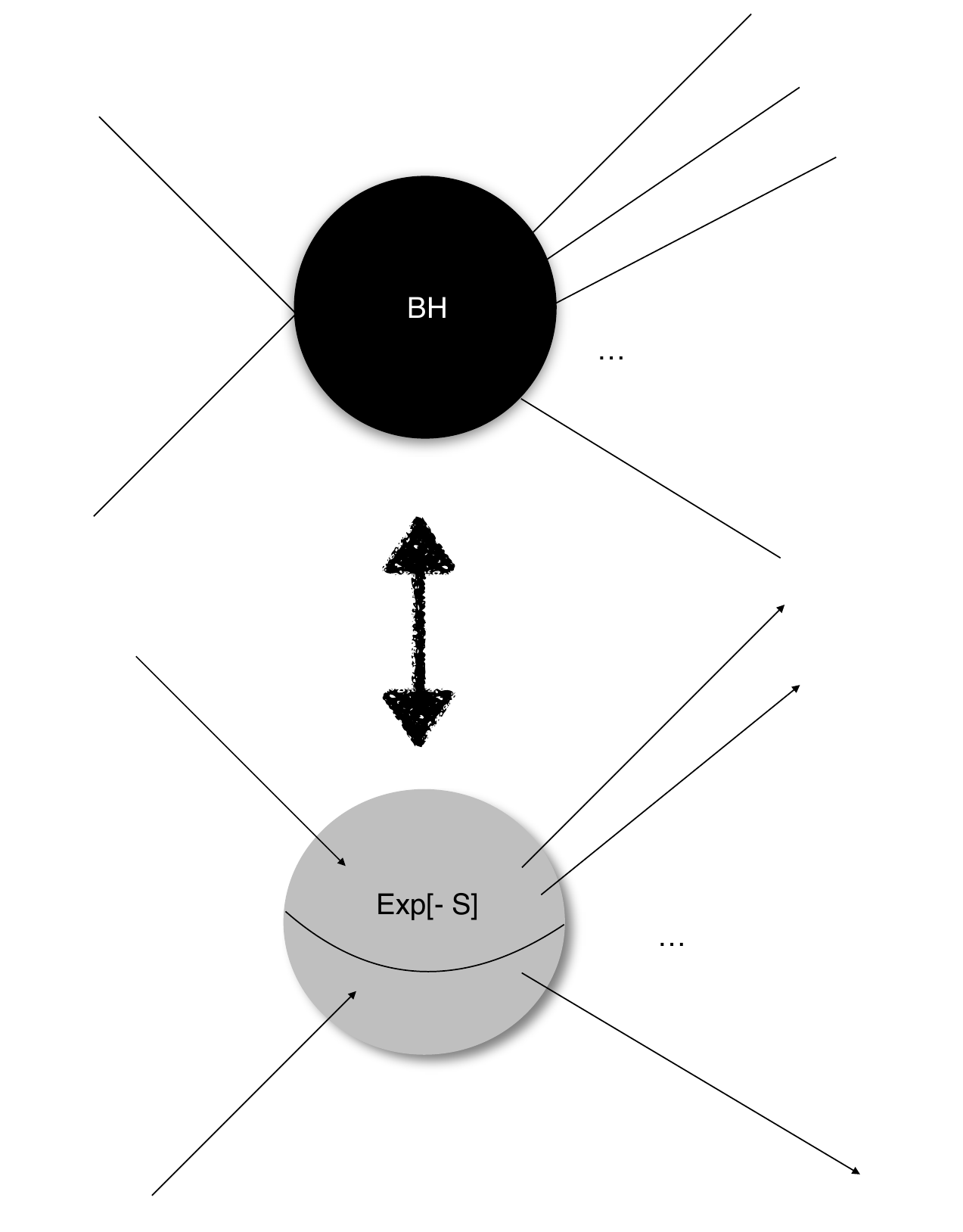}}
\caption{The inclusion of a BH threshold in gravitational scattering amplitudes is equivalent to a de-localization of the interaction vertex with an entropic form factor ${\rm Exp}[-S]\equiv {\rm Exp}[-L_{Pl}^{2}\Box]$.    }
\end{figure}

The main idea is that sub-planckian lengths cannot be probed from scatterings with a CM energy which is higher than the Planck energy:
this may correspond to the creation of many soft quanta rather than on probing the UV sub-planckian modes. 
Such a phenomena is what {\it Banks} defined {\it Asymptotic Darkness} while {\it Dvali} and {\it Gomez} as {\it Classicalization}. 
Although this idea is certainly attractive, I think no any convincing argument so far was suggested, as a definitive theoretical evidence of such a dynamics \footnote{See also Refs. 
\cite{Amati:2007ak,Ciafaloni:2015vsa,Dvali:2014ila,Kuhnel:2014xga,Addazi:2016ksu,Ciafaloni:2017ort} for other previous attempts to explore the unitarization of gravity in Transplanckian scattering amplitudes.}. 
However, we do not have any clear no-go proofs against self-unitarization. 
This renders the situation puzzling and interesting!
The purpose of this letter is to share some considerations on the BH formation in quantum gravity. 
We will see that the Banks-Dvali-Gomez (BDG) proposal is surprisingly related to the emergence of non-locality in quantum gravity. 
This may provide a potentially strong breakthrough towards our understanding of the UV quantum gravity behavior
as a dynamical self-superrenormalizable theory. 
First of all, we have an apparent problem between the Heisenberg's and BH radius laws: 
\begin{equation}
\label{HUP}
 \Delta R\, \Delta P_{R}\geq \frac{\hbar}{2} 
\end{equation}
$${\bf Vs}$$
\begin{equation}
\label{AHUP}
\Delta R\sim 2G_{N}\Delta P_{R}\, ,
\end{equation}
where $R$ is the radial coordinate from the BH center (also identified with the impact parameter in Gravitational Scatterings).
If the cross-section, as in Eq.\ref{cro}, quadratically grows with the CM energy and the BH radial distance,
then this seems to be in an apparent contradiction with the Heisenberg's uncertainty principle.
Thus we seem to find a paradox: the BH geometrical cross-section seems to not be fully compatible with the HUP,
when we consider the zero resolution limit.
In a BH formation, energetic scales seem to fix its geometric length scales. 
In other words, the BH seems to live on a state that seems to not satisfy the HUP in any resolution limits.
Thus, this is the situation in a nutshell: when energies are below the threshold of the BH production, 
perturbative quantum gravity on a flat background follows the HUP and, therefore, length resolutions scale as the inverse of energy ones;
but when the BH is produced, HUP is violated and energy resolutions start to scale as the lengths. 

Insightfully, combining Eq.\ref{HUP} with Eq.\ref{AHUP}, we obtain the following bound on the 
radial coordinates as (numeral constant omitted)
\begin{equation}
\label{Dell}
\Delta R\geq l_{Pl}\, .
\end{equation}
This seems to imply that the two principles can be self-consistently implemented if and only if the 
accessing to any sub-planckian lengths is prohibited. 

Let us focalize on the main consequence of the growing of the 
BH geometrical cross-section: the field vertices interpolating for a BH production 
cannot be thought as local anymore, because of the Eq.\ref{AHUP}.
This implies that the entire quantum field theory approach based on Fourier transforms and Feynman's diagrams 
becomes obsolete and it has to be rethought renouncing to the HUP based particle-wave duality. 
Fortunately, the BH production case can be understood from conventional perturbative field theory approaches 
if we delocalize the interaction vertices with a form factor as 
\begin{equation}
\label{FF}
\mathcal{F}\sim e^{-R_{S}^{2}/L_{Pl}^{2}}\sim e^{-L_{Pl}^{2}E^{2}}\sim  e^{-L_{Pl}^{2}\Box}\, . 
\end{equation}

The why this form factor should have such an exponential profile will be clear from the following discussions. 
For the moment, we just point out the evidence that any perturbative quantum gravity vertices must be smeared with 
a non-local factor as Eq.\ref{FF}; as an effective inclusion of the BH production threshold. In this sense, the BH production introduces
the violation of the locality principle. The BH formation/evaporation is interpreted as a non-local collective phenomena. 
But, at this point of our arguments, still, how, this new law can emerge out, may sound mysterious.
We will see that Eq.\ref{FF} has an entropic origin related to the holographic memory of BHs.
On the other hand, Eq.\ref{FF} implements the very same condition of Eq.\ref{Dell}, self-consistently. 

For the next step, the key point is that the BH cross-section scales as its area, in turn holographically storing all BH informations.
This suggests that the main point for comprehending the apparently paradoxical Eq. \ref{AHUP} consists in considering the entropic 
and thermodynamical aspects of the process. Indeed, the BH production is related to the inclusion of an entropic factor 
that is large as the BH size as 
\begin{equation}
\label{BH}
\sigma_{G}\sim S_{BH}\, ,
\end{equation}
where $S_{BH}$ is the BH entropy. 
Thus, this means that more we increase the energy, more we increase the geometric cross-section, more we increase the BH size, more we increase the entropy!

Now, we wish to interpret the entropic content of the produced BH as dominated by {\it dynamical quantum hairs} 
\cite{V1,Coleman:1991ku,Preskill:1992tc,Giddings:1993de,Dvali:2012rt,V2} 
or {\it hairons} \cite{Addazi:2020vhq,Addazi:2020hrs}:
\begin{equation}
\label{sigmanNN}
\sigma_{G}\sim S_{BH}\sim N\, . 
\end{equation}
The BH formation means a creation of a certain amount of new information stored in the new $N$ hairons. 
We will see that this is exactly why the HUP is violated around the BH threshold:
as pointed out in our recent compainion works, the BH holographic entropy will induce a new form 
of gravitational decoherence in the system  \cite{Addazi:2020vhq,Addazi:2020hrs}.
This leads to reconsider the graviton-graviton interaction vertices:
when the BH is formed, any interaction vertices 
are associated to the production of entropy, in turn related to the generation of quantum information.
In other words, quantum gravity amplitudes are unitarized as an entropic protection mechanism.

When the BH threshold is achieved, the generic $2\rightarrow 2$ scattering process
is reinterpreted as the following sequential transition amplitudes: 
\begin{equation}
\label{twotwo}
\mathcal{S}(2\rightarrow 2)=\langle 2_{in}|BH\rangle \langle BH|\mathcal{S}|BH\rangle \langle BH|2_{out}\rangle\, .
\end{equation}
On the other hand, the BH information storage leads to 
the identification of the BH state as
\begin{equation}
\label{BH}
|BH\rangle=|T_{BH}\rangle=|N\rangle\, ,
\end{equation}
where N is the number of hairons, storing N-qubits, and $T_{BH}$ is the hairon average energy interpreted as the BH temperature. 
Now, from the information theory prospective, the $\langle 2_{in}|N\rangle$ transition corresponds to the 
generation of N-qubits from the initial information carried by the in-states. 
On the other hand, the $\langle N|2_{out}\rangle$ corresponds to the N-bits annihilation into few ones carried by the out state. 
In the large N-limit, both the amplitudes have a high entropy cost as 
\begin{equation}
\label{NN}
\langle 2_{in}|N\rangle\simeq \langle N|2_{out}\rangle\simeq e^{-S_{BH}}\sim e^{-N}\,. 
\end{equation}
Such a transition gives to the form factor, suggested in Eq.\ref{FF}, an entropic meaning as 
\begin{equation}
\label{FFF}
\mathcal{F}\sim e^{-S_{BH}}\sim e^{-L_{Pl}^{2}\Box}\, .
\end{equation}
On the contrary, the probability that a BH remains to the same state, in the large N limit 
is 
\begin{equation}
\label{BHBH}
\langle BH|\mathcal{S}|BH\rangle \sim 1-e^{-N}
\end{equation}
(omitting every constant prefactors not relevant for our discussion aims). Indeed, even if the $\mathcal{S}$-operator is evaluated on
two not asymptotical states, a BH with a large amount of information is in an almost stable state, rendering its evaluation 
exponentially close to the realistic process. 

Therefore, integrating on the phase space, from the Eq.\ref{twotwo}, the geometric cross-section is unitarized by the entropic form factor,
in the large N limit, as
\begin{equation}
\label{twotwo}
\mathcal{S}\rightarrow \sigma_{G,N}\sim e^{-N}N\rightarrow e^{-L_{Pl}^{2}s}G_{N}s\sim e^{-L_{Pl}^{-2}R_{BH}^{2}}R_{BH}^{2}\, . 
\end{equation}

This argument can be generalized to a generic $2\rightarrow M$ process. 
If all quanta are produced with the same energy, we will obtain the following exponentialization: 
\begin{equation}
\label{twotwo}
|\mathcal{S}(2\rightarrow M)|^{2}\sim e^{-N+M}\, .
\end{equation}

Therefore, the dominating cross-section corresponds to $N\simeq M$, saturating the unitarity bound \footnote{Such a result is compatible with the soft limit, when the BH is not formed yet, obtained from the application of Weinberg's soft theorems on the tree-level high multiplicity cross-section, as shown by myself, {\it Bianchi} and {\it Veneziano} in Ref.\cite{Addazi:2016ksu}.}.
This was somehow expected, since the BH formation/evaporation process is thought as
a high multiplicity scattering. Indeed the BH will evaporate out to Hawking radiation which is imagined carrying 
the large amount of quantum hairs stored in the horizon. 

The next step is to show that such a dynamics is stable against radiative corrections, 
in self-consistency with the $\{\Delta t\sim \Delta E\}$ BH law. 

\vspace{0.1cm}

{\it Perturbation theory on a curved background}. Let us consider the gravitational lagrangian,
linearized around the background metric $\bar{g}_{\mu\nu}$ as
 $g_{\mu\nu}=\bar{g}_{\mu\nu}+h_{\mu\nu}$ (see the Appendix for more details).

Let us consider a generic source particle fields propagating in the space-time background. 
The field provides an energy momentum tensor, which is the gravitational source $J_{\mu\nu}\equiv T_{\mu\nu}$. 
It is commonly retained that the gravitational interaction can be described as
\begin{equation}
\label{JDJ}
J_{\mu\nu}D^{\mu\nu}_{\rho\sigma}J^{\rho\sigma}\, , 
\end{equation}
where $D$ is the graviton propagator interpolating between the two sources. 
Indeed, the partition function is 
\begin{equation}
\label{Z}
Z=\int \mathcal{D}h_{\mu\nu}e^{i\int d^{4}x\Big[ -\frac{1}{2}(h D h) +(J h) \Big]}\, , 
\end{equation}
where $(,)$ omits well understood tensor contractions. 

The corresponding generating functional reads as 
\begin{equation}
\label{Zd}
Z(J)=C e^{-\frac{i}{2}\int \int d^{4}x d^{4}y (J(x) D(x-y) J(y))}=Ce^{iW(J)}\, . 
\end{equation}

In the following, we will show how the partition function is effectively modified by the BH threshold. 
To arrive to it, 
let us return to a S-matrix scattering $in \rightarrow out$
on a curved space-time background. 
The $in$ and $out$ particles cannot be considered as completely disentangled by 
quantum hairs stored in any space-time background.  
The holographic entanglement scrambling time is 
$$t_{scr}\sim R\, {\rm log}S\sim \sqrt{N}\log(N)\, t_{Pl}\, .$$
where $R\sim \sqrt{N}l_{Pl}$ (see Refs. \cite{Hayden:2007cs,Sekino:2008he,Shenker:2013pqa}).

 Therefore we introduce the entangled states  
\begin{equation}
\label{enn}
|N\rangle \otimes | in\rangle\simeq e^{-N}|in\rangle,\,\,\,\ |out\rangle  \otimes |M\rangle =e^{-M}|out\rangle \, ,
\end{equation}
where $N,M$ are the number of hairs populating the vacuum state. 
 For $N,M>>\#_{in,out}$, 
one can safely assume that $N\simeq M$.
Therefore, the bare S-matrix is dressed as 
\begin{equation}
\label{S}
\mathcal{S}=\langle in | \otimes\langle N|\mathcal{S}|M\rangle \otimes |out\rangle\simeq e^{-N}\mathcal{S}_{0}\, ,
\end{equation}
$$S_{0}=\langle in_{flat}|\mathcal{S}|out_{flat}\rangle\,  .$$

Let us consider the example of a $2\rightarrow m$ scattering. 
From Eq.\ref{S} leads to 
\begin{equation}
\label{S}
\mathcal{S}=\langle 1,2 | \otimes\langle N|\mathcal{S}|M\rangle \otimes |1',....m'\rangle\simeq e^{-N} \mathcal{S}_{0}(2\rightarrow m)\, .
\end{equation}

Even if $\mathcal{S}_{0}$ would polynomially or exponentially diverge with energy, the environment 
would suppress the process if the final multiplicity of the $2\rightarrow m$ process 
departs from the number of hairs.

This space-time dressing factors percolate into a re-definition of the current sources in Eqs.[\ref{Z},\ref{Zd}] (see Fig.1):
\begin{equation} 
\label{J}
J\rightarrow e^{-N}J\rightarrow e^{-L_{Pl}^{2}\Box}J\, . 
\end{equation}
Eq.\ref{J} leads to an effective smearing of the interaction vertices
with the entropic form factor. 
Indeed, this can be dually seen as a smearing of the graviton propagator \footnote{Previous 
explorations of the formation of non-perturbative classicalon and non-locality can be found in our 
previous work   \cite{Addazi:2015ppa} in the case of non-local scalar field theory.}
as
\begin{equation}
\label{Dmunu}
D_{\mu\nu}\langle T\{(h(x)h(y))_{\mu\nu}\}\rangle\rightarrow d_{\mu\nu}e^{-L_{Pl}^{2}\Box}\frac{1}{\Box}\, . 
\end{equation}

Thus, this is a signal that quantum gravity is a super-renormalizable theory 
if including the non-perturbative sector.
In this sense, quantum gravity is secretly a UV complete theory without introducing any 
new heavy degrees of freedom into it. On the other hand, super-renormalizability 
emerges out without introducing any explicit non-local actions 
as proposed in Refs.
\cite{NL1,NL2,NL3,Modesto:2011kw,Modesto1,Modesto2,Modesto3,Biswas:2011ar,Mazumdar} (see also Ref.\cite{Briscese:2019twl} for recent discussions on UV asymptotic safety properties of 
non-local quantum gravity
 \footnote{The Wilsonian renormalization approach applied to quantum gravity provides for other non-local terms, which  do not improve 
 the UV behaviour (see Refs. \cite{Buchbinder:1992rb,Elizalde:1995tx}).}.

Therefore, quantum gravity is intrinsically non-local even if this is hidden in the Einstein-Hilbert action. 
Indeed, this may be an explanation of why it is not possible to define any local energy-momentum tensor for the gravitational field itself. 

On the other hand, the Eqs.[\ref{HUP},\ref{AHUP}] can be effectively interpolated by 
an effective Generalized Uncertainty Principle GUP \cite{KM,ACV1,ACV2,ACV3} as
\begin{equation}
\label{GUP}
 \Delta R\Delta P_{R}\geq \frac{\hbar}{2}(1+\frac{1}{M_{Pl}^{2}}\Delta P_{R}^{2})+...
\end{equation}

In the UV limit $\Delta P>>M_{Pl}^{2}$, this converges to 
\begin{equation}
\label{CLl}
\Delta X_{R} \geq \Big(\frac{\hbar L_{Pl}}{2M_{Pl}}\Big)\Delta P_{R}\, , 
\end{equation}
where the BH saturates the equality bound. This prohibits the UV accessing, also avoiding for any causality issues -- usually problematic 
for any introduction of non-local quantum operators by hand. 


\begin{figure}[ht]
\centerline{ \includegraphics [width=1.1\columnwidth]{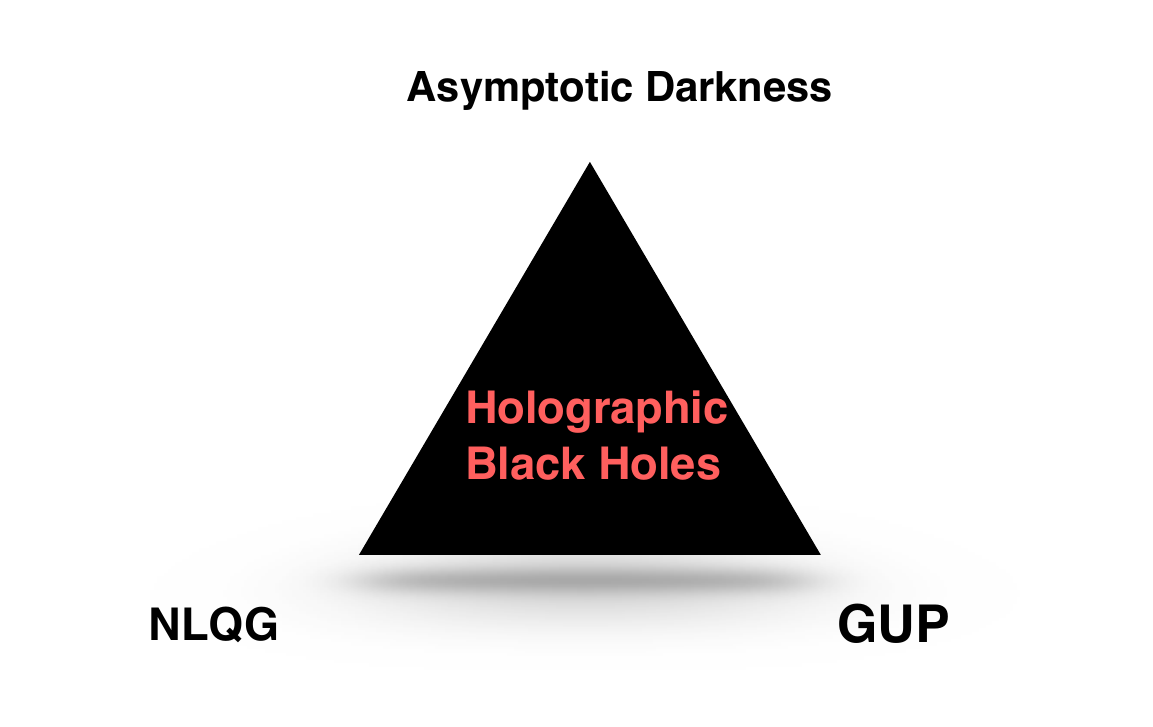}}
\caption{  The inclusion of Black Holes, with a holographic entropy, into scattering amplitudes, leads to a series of dualities among the Non-local Quantum Gravity (NLQG), the Generalized Uncertainty Principle (GUP)
and Asymptotic Darkness. All these approaches reflect the same dynamical unitarization of transplanckian gravitational scatterings.  }
\end{figure}



\vspace{0.1cm}

{\it Conclusions and remarks}. In this letter, we have shown that the inclusion of the Black Hole
in gravitational scattering amplitudes will unitarize any possible quantum gravity processes.
Our arguments  sustain the Banks-Dvali-Gomez (BDG) conjecture that 
quantum gravity is a self-unitarizing theory. 
We have seen that the self-completion dynamics is related to 
a peculiar aspect of Black Holes: the holographic information storage.
Thus, when the BH threshold is achieved, the entropic protection will save quantum gravity 
from any non-unitarity disasters. The entropic barrier prohibits for any processes that 
arbitrary create and destroy a large amount of quantum information in the Universe  \cite{Addazi:2020vhq,Addazi:2020hrs} \footnote{From the information theory prospective, this seems to work pretty similarly to 
topological quantum computing, where typically the high quantum degeneracy of the ground state 
provides for a non-local protection of qu-bits from any noise and fluctuations \cite{Nayak:2008zza}.
In our related paper, in preparation, we will show that such an analogy may be more than just a case:
 the BH memory can be related to the topological complexity of the space-time \cite{Addazi:2020mnm}.The super-planckian localization of energy in a BH radius may be understood as 
a phase transition jump of the gravitational susceptibility $\langle R\tilde{R}\rangle$ in {\it vacuo} \cite{Addazi:2020mnm}. In this sense, the BH production can be re-thought as produced at a quantum critical temperature
crossing a topological phase transition. }.
On the other hand, the Heisenberg's uncertainty principle (HUP) flows to a dynamical break down
caused by non-perturbative effects. 
Thus, the Black Hole formation/evaporation processes can self-unitarize quantum gravity,
subtly unveiling hidden non-locallity of quantum gravity.
This process seems to render quantum gravity as UV insensitive to any sub-planckian wave length modes. 



\vspace{0.2cm}

{\it Acknowledgments}. I wish to thank Fabio Briscese and Leonardo Modesto for discussions and remarks on these aspects.

\onecolumngrid


\twocolumngrid

\end{document}